\documentclass[12pt,a4paper]{article}
\usepackage{amssymb}

\newtheorem{thm}{Theorem}
\newtheorem{col}{Corollary}
\newtheorem{lem}{Lemma}

\newenvironment{proof}{\begin{trivlist} \item[] {\em Proof}. }%
{\hfill $\square$ \end{trivlist}}

\title{
INTEGRABLE SYSTEMS WITH PAIRWISE INTERACTIONS AND FUNCTIONAL EQUATIONS\\}

\author{H. W. Braden\\
\normalsize
\em Department of Mathematics and Statistics, University of Edinburgh, \\
\normalsize
\em Edinburgh, UK. \\
\\
V. M. Buchstaber\\
\normalsize
   \em      National Scientific and Research Institute of\\
\normalsize
     \em    Physico-Technical and Radio-Technical Measurement,\\
\normalsize
    \em     VNIIFTRI, Mendeleevo, Moscow Region\\
\normalsize
      \em   141570, Russia\\
}

\begin{document}

\begin{titlepage}
\renewcommand{\thepage}{}
\maketitle

\begin{abstract}
A new ansatz  is presented for a Lax pair describing
systems of particles on the line interacting via
(possibly nonsymmetric) pairwise forces. Particular cases of this yield
the known Lax pairs for the Calogero-Moser and Toda systems, as well
as their relativistic generalisations. The ansatz leads to a system
of functional equations.  Several new functional equations
are described and the general analytic solution to some  of these is given.
New integrable systems are described.
\end{abstract}
\vfill
\end{titlepage}
\renewcommand{\thepage}{\arabic{page}}

\section{Introduction}
Completely integrable systems arise in various diverse settings in
both mathematics and physics and accordingly have been
studied from many different points of view, 
a fact which underlies their importance and interest
(see for example \cite{Nov}.)
Within this area the study of Lax pairs, a zero curvature
condition, plays an important role.
The construction of such Lax pairs has followed many routes and
this paper will further investigate the connection between functional
equations and such zero curvature conditions.
The essential idea in this approach is to reduce the constraints
of the Lax pair ${L,M}$ implicit in ${\dot L}= [L,M]$
to that of a functional equation. Our study
will broaden the ansatz for the Lax pair and correspondingly lead to
a more general functional equation than hitherto studied.
This enables us to understand many of the known integrable systems
(and their corresponding functional equations) from a unified
perspective. The symmetries
of the functional equation we obtain are very large and this group
relates distinct functional equations and their corresponding physical
models. We feel this connection between functional equations and
completely integrable systems is part of a broader and less well
understood aspect of the subject. 
Functional equations have of course a long and interesting history
in connection with mathematical physics and touch upon many branches
of mathematics \cite{Acza, Acz}.
Novikov's school for example considered the Hirzebruch genera associated
with the index theorems of known elliptic operators and showed that these 
arose as solutions of functional equations.
More recently Ochanine showed the string inspired Witten index could
be described by Hirzebruch's construction where the the functional equation
was that appropriate to an elliptic function.
These same functional equations arise (as we shall later see in more
detail) in the
context of completely integrable systems.
The latter also appear in the study of conformal and string theories
and this  connection between string
theory and  finite dimensional completely integrable systems needs to be
better understood.
The functional equations and integrable systems we shall discuss
arise naturally in investigations of the KP and KdV equations
\cite{AMM, Kr1}.

To make matters
concrete let us consider how such functional equations arise in the
context of integrable systems of particles on the line.
Here one starts with an ansatz for the matrices $L$ and $M$ of the Lax pair
and seeks the restrictions  necessary to obtain
equations of motion of  some desired
form. These restrictions typically involve the study of functional equations.
The Calogero-Moser \cite{Ca}
system provides the paradigm for this approach.
Beginning with the ansatz (for $n\times n$ matrices)
$$L_{jk}=p_j\delta_{jk}+ \, g\,(1-\delta_{jk}) A(q_j-q_k),\quad
  M_{jk}=g\, [\delta_{jk}\sum_{l\ne j}B(q_j-q_l)-(1-\delta_{jk})C(q_j-q_k) ]
$$
one finds ${\dot L}= [L,M]$
yields the
equations of motion for the Hamiltonian system ($n\ge3$)
$$H={1\over 2}\sum_{j}p_j\sp2 +g\sp2\sum_{j<k}U(q_j-q_k)
\quad\quad\quad U(x)=A(x) A(-x) + {\rm constant}$$
provided $C(x)=-A\sp\prime(x)$,
and that $A(x)$ and $B(x)$ satisfy the functional equation
\begin{equation}
A(x) A\sp\prime(y) -A(y)A\sp\prime(x)=A(x+y)[ B(x)-B(y)].
\label{calfun}
\end{equation}
The solutions to this functional equation may be expressed in terms
of elliptic functions and their degenerations.\footnote{
The solution to this equation has been given by various authors
with assumptions of even/oddness on the functions appearing
\cite{Ca2, OPc, PS} or assumptions on the nature of $B$ \cite{Kr1}.
The general solution was given in \cite{BCb, Bu1}.
The derivation we shall present later in fact yields the even/oddness
assumptions of these earlier works.}
Krichever used this functional equation in his proof
of the \lq rigidity\rq\ property of elliptic genera \cite{Kr}
and it appears when discussing rational and pole solutions
of the KP and KdV equations \cite{Kr1, AMM}.
Different  starting ansatz lead to the relativistic \cite{RS,BCa,Ra}
Calogero-Moser systems, the Toda \cite{To,OPa}
and relativistic Toda \cite{BRa,BRb}
equations. Underlying the corresponding functional equations of each
of these models lies the addition formula for elliptic functions.
Further,  in studying the quantum mechanics of these systems
similar functional equations arise when factorising the ground
state wave function \cite{BP}.

In this paper we will introduce a new ansatz
that includes the previous examples (together with their functional
equations) as special cases. 
We shall be seeking Lax pairs that lead to equations of the form
\begin{equation}
{\ddot q_j}=\sum_{k\ne j}(a+b \dot q_j) (a+b \dot q_k) 
V_{jk}(q_j-q_k).\label{ans} 
\end{equation}
Here we are allowing the interaction\footnote{
When $b=0$ then $V_{jk}$ is just $-{d\over dq_j}U(q)$, where $U(q)$ is the
potential energy of the system. When $b=1$ and $a=0$ then the Hamiltonian
for the system is \cite{RS} $H=\sum_j \cosh\theta_j \ V_j(q)$ where
$\theta_j$ is the rapidity canonically conjugate to $q_j$ and
$V_j(q)=\prod_{k\ne j} V_{jk}$. In this case (\ref{ans}) corresponds to the
flows of $S_{\pm}=\sum_j e\sp{\pm\theta_j} \ V_j(q)$.}
$V_{jk}$ to in principle depend
on the choice of pair ${j,k}$: when $V_{jk}$ is the same for all pairs
we have a system of Calogero-Moser type while if $V_{jk}$ is the same for
pairs ${j,j\pm 1}$ and zero otherwise we have a Toda system (of $A_n$ type).
When the constant $b=0$ we obtain the nonrelativistic systems and when
$a=0$ we obtain the relativistic systems previously examined.
Indeed,  because the interactions of the
system only depend on coordinate differences, the shift 
$q\rightarrow q-a\tau\, t/b$ enables us without any loss of generality
to set $a=0$  when $b\ne0$ and we shall do this where appropriate.
Our first stage of generalisation
then is to consider a matrix of pairwise (though in principle
distinct) interactions
and it is just this limitation to pairwise interactions $V_{jk}=V_{jk}(q_j-q_k)$
that enables us to derive functional equations of a given type.
Yet rather than just a single functional equation our extension allowing
different interactions $V_{jk}$ now leads to a {\em system} of functional equations,
and this interplay of matrix relations and functional equations appears new.
Although our ansatz by its very form must include the  Calogero-Moser and Toda models,
our approach shows how they may be unified by the study of one functional
equation.

We find the functional equations  needed to construct a Lax pair
yielding the equations of motion (\ref{ans}) are of the form (for $b\ne0$)
\begin{equation}
\phi_1(x+y)=
{ { \biggl|\matrix{\phi_2(x)&\phi_2(y)\cr\phi_3(x)&\phi_3(y)\cr}\biggr|} 
\over 
{ \biggl|\matrix{\phi_4(x)&\phi_4(y)\cr\phi_5(x)&\phi_5(y)\cr}\biggr|} }
\label{functional}
\end{equation}
Elsewhere  we have shown,
\begin{thm}{\cite{BB}}
The general analytic solution to the
functional equation (\ref{functional})\ is, up to a ${\cal{G}}$ action given
by (\ref{symm}-\ref{symmsb}), of the form
$$\phi_1(x)= {\Phi(x;\nu)\over\Phi(x;\mu)},\quad
  {\phi_2(x)\choose\phi_3(x)}={\Phi(x;\nu)\choose\Phi(x;\nu)\sp\prime}\quad
{ and}\quad
  {\phi_4(x)\choose\phi_5(x)}={\Phi(x;\mu)\choose\Phi(x;\mu)\sp\prime}.
$$
Here
$$\Phi(x;\nu)\equiv {\sigma(\nu-x)\over {\sigma(\nu)\sigma(x)}}\,
  e\sp{\zeta(\nu)x}
$$
where $\sigma(x)=\sigma(x|\omega,\omega\sp\prime)$ and
$\zeta(x) ={\sigma(x)\sp\prime \over\sigma(x)}$
are the Weierstrass sigma and zeta functions.
\end{thm}
The symmetries ${\cal G}$ of (\ref{functional})\ will
be described in section three. The proof given in \cite{BB} is constructive and the
transformations needed to obtain the solutions may be readily implemented. 

The case $b=0$ is  more problematical.  In this case we
obtain functional equations of the form
\begin{equation}
\phi_6(x+y)=\phi_1(x+y)\big( \phi_4(x)-\phi_5(y)\big)
+{ \biggl|\matrix{\phi_2(x)&\phi_3(y)\cr\phi\sp\prime_2(x)&\phi\sp\prime_3(y)
\cr}\biggr|}.
\label{functional2}
\end{equation}
Certainly we may
take the limit $b\rightarrow0$ to our solutions of (\ref{functional})
(which, for example, will give the Calogero-Moser model as the
nonrelativistic limit of the  relativistic Calogero-Moser model)
to obtain solutions of (\ref{functional2}) but at present we don't
know the general solution to (\ref{functional2}). We can show however
that known nonrelativistic models are solutions to this equation,
together with new potentials such as
$$
V_{jk}(x)=a_j a_k\,  \wp\sp\prime(x).
\label{eq:newnonrel}
$$
When $a_j=a_k$ this yields the usual type IV
Calogero-Moser potential.

We remark that (\ref{functional}) and a suitably symmetrized
form of (\ref{functional2}) are
particular cases of the functional equation
\begin{equation}
\sum_{i=0}\sp{N}
\phi_{3i}(x+y)
 { \biggl|\matrix{\phi_{3i+1}(x)&\phi_{3i+1}(y)\cr
                  \phi_{3i+2}(x)&\phi_{3i+2}(y)\cr}\biggr|}=0
\label{Bfun}
\end{equation}
with $N=1$ in the case $b\ne0$ and $N=2$ in case $b=0$.
In the case $\phi_{3i+2}=\phi_{3i+1}\sp\prime$
Buchstaber and Krichever have discussed  (\ref{Bfun}) in connection with
functional equations satisfied by Baker-Akhiezer functions \cite{BK}.
Dubrovin, Fokas and Santini \cite{DFS} have also investigated
integrable functional equations via algebraic geometry.

A further generalisation of the  Calogero-Moser system has been to
associate such an integrable system to the
root system of an arbitrary semisimple Lie algebra \cite{OPa,OPb}.
At this stage of generalisation we have essentially the Lax pairs
associated with $A_n$ type root systems. To incorporate more general
root systems we may consider
embedding $\Omega :R{\sp n}\rightarrow R{\sp N}$ of our $n$-degrees of
freedom into a larger space with the interactions $V$
still of the given form. We will not present this generalisation
here.

This paper then  presents an ansatz for Lax pairs whose
consistency yields equations of motion (\ref{ans}) and
the corresponding functional equations (\ref{functional}) and 
(\ref{functional2}). We shall show how various specialisations
lead to the known systems and introduce some new ones. 
An outline of the paper is as follows.
In section two we present the ansatz.
For clarity of exposition we initially confine our attention to the case
$b\ne0$ returning to the $b=0$ case in Section five. Here we
determine in Theorem $2$ the system of functional equations to be solved, reducing
the nontrivial equations to be solved to the form (\ref{functional}). 
Section three describes the invariances of (\ref{functional}) and
illustrates the solution of Theorem $1$ as a means of introducing certain 
elliptic function identities useful in the sequel.
Here we will apply the 
general analytic solution of this  theorem to our
system of equations (Corollary $1$) and then  as an example
show how the relativistic example of Bruschi and Calogero  \cite{BCa}
arises.
In general our ansatz leads to a system of functional equations
and Section four
looks at the constraints imposed on the parameters of the solution
to (\ref{functional}) by such a system. Theorem $3$ determines these
constraints and these are illustrated by  the relativistic 
Calogero-Moser and Toda models. Further, we are able to characterise the
relativistic Calogero-Moser model by a certain \lq generic\rq\ property,
Theorem $4$.
Section five returns to the $b=0$ case deriving the appropriate
functional equation. Again, several examples will be given.
Finally we conclude with a brief discussion.

In an earlier version of this paper we proved Theorem 1.
Subsequently we found a direct an constructive proof which has been
preseted separately \cite{BB}. In revising the present paper accordingly
we have also strengthened the results of section $4$.

\section{The Ansatz}
We shall now describe the ansatz, 
introducing our notation
and illustrating  some techniques useful in the
reduction problem of Lax pairs to functional equations.
Having presented the ansatz for our Lax pair we proceed to determine
the restrictions  on the functions that appear in this.
The equations we find are a natural generalisation of those found
in  \cite{BCa}.
We will then seek the relevant functional equations to be solved in later
sections.

We need a few definitions in order to specify our ansatz.
Let $\tau$ be the fixed vector $\tau=(1,\ldots,1)$ and denote 
by $X_d={\rm Diag}(X_1,\ldots,X_n)$ the
injection ${\mathbb R}{\sp n}\rightarrow Mat(n)$. 
Further let 
${\cal M}_n=\{A | A\in\, Mat(n),\, A_{ii}=0,\, A_{ij}=A_{ij}(q_i-q_j)\}$
be a subset of matrix-valued functions of one variable.
(One can also extend our analysis to
the case of nonvanishing diagonal elements.) 
Note this set depends on a choice of coordinates
$\{e_i\}$ with respect to which we
express our matrices and determine the coordinate projection
$q_i=(e_i,q)$.
A change of basis results in a straightforward conjugation.
We denote by $e_{ j}=(0,\ldots,{1},\ldots,0)\sp T$ the  $j-$th
coordinate (column) vector.
Thus ${e_j\sp T}.e_k=\delta_{jk}$, ${e_j\sp T}.A.e_k=A_{jk}$
and $X_d.e_k=X_k e_k$. 

Our ansatz for the Lax pair takes the form\footnote{
We remark that there is no essential change if we take
$$L(q)=\dot q_d +(a \tau+b \dot q)_d\sp\epsilon\, A 
(a \tau+b \dot q)_d\sp{1-\epsilon}
$$
and a similar modification to $M$, for any value of $ \epsilon$.
We choose the symmetrical value given.}
\begin{eqnarray}
L(q)&=&\dot q_d +\sqrt{(a \tau+b \dot q)_d}\, A \sqrt{(a \tau+b \dot q)_d}\cr
M(q)\,&=&
(B.[a \tau+b \dot q])_d 
+\sqrt{(a \tau+b \dot q)_d}\, C\sqrt{(a \tau+b \dot q)_d}
\label{ansatz}
\end{eqnarray}
where $A,B,C\in{\cal M}_n$. 
When $a=0$ the matrix $\sqrt{\dot q_d}$ corresponds to the diagonal
matrix $D$ of \cite{Rb}.
This  Lax operator also  possesses the symmetry 
\begin{equation}
L\big(q,b,a,A(q)\big)=l\ L\big(q/l,bl,a, l\sp{-1} A(l\cdot q/l)\big),
\label{eq:Laxsymm}
\end{equation}
which enables us to  rescale $b\ne0$ to $1$ and we shall later do this.
It will be convenient to define 
\begin{equation}
G\equiv{1\over b}{\psi\sp\prime \over\psi}= B+{b\over2}V\quad\quad{\rm and}
\quad\quad H=A\sp\prime-C.
\label{eq:defgh}
\end{equation}
(This defines $\psi$ up to a multiplicative factor that is immaterial in what follows.)
Then we have
\begin{thm}
When $b\ne0$ the Lax pair (\ref{ansatz}) yields the equations of motion (\ref{ans})
if and only if
\begin{eqnarray}
-H_{jk} + b A_{jk}G_{jk}&=&0,\label{solb}\\
H_{jk}- b A_{jk}G_{kj} +b\sp 2 A_{jk}V_{kj}&=&0,\label{sola}\\
b\, A_{jk}(G_{jm}-G_{km})+b\sp2 A_{jk} V_{km}&=&
  b\, \biggl|\matrix{A_{jm}&A_{mk}\cr C_{jm}&C_{mk}\cr}\biggr|.
\label{solc}
\end{eqnarray}
with
\begin{equation}
V_{jk}={1\over {1-b\sp2 A_{jk}A_{kj}}}
     \biggl|\matrix{A_{jk}&A_{kj}\cr A\sp\prime_{jk}&A\sp\prime_{kj}\cr}\biggr|
={1\over b\sp2}{d\over dx}\ln(1-b\sp2 A_{jk}(x)A_{kj}(-x)).
\label{eq:vjk}
\end{equation}
In particular,  $V_{jk}\ne0\Rightarrow A_{jk},\ A_{kj}\ne0$.
\end{thm}

Observe that once we have established the theorem an ansatz is entirely
specified by the matrices $A$ and $G$: given $A$ and $G$ then $H$ is determined
by (\ref{solb}) and so $C$ by (\ref{eq:defgh}); also $V$, determined via
$A$ using (\ref{eq:vjk}), together with $G$ are sufficient to determine $B$. 
Thus our goal will be
to find such matrices $A$ and $G$. We remark in passing that a
consequence of (\ref{eq:vjk}) is that $V_{jk}(x)=-V_{kj}(-x)$
and so $\sum \dot q_j$ is constant. We begin by establishing the theorem.
\begin{proof}
By using the freedom to commute diagonal matrices we observe that
$${\dot L}= \sqrt{(a \tau+b \dot q)_d}\,\biggl(
  {{\ddot q}_d\over (a \tau+b \dot q)_d} + {b\over2}\Bigl(
  {{\ddot q}_d\over (a \tau+b \dot q)_d} A +
 A{{\ddot q}_d\over (a \tau+b \dot q)_d}\Bigr)
  +[{ \dot q}_d,A\sp\prime ]\biggr) \sqrt{(a \tau+b \dot q)_d}.
$$
Here ${{dA}\over{dt}}=({{dA_{ij}}\over{dt}})=([{\dot q}_i-{\dot q}_j]
A\sp\prime_{ij})=[\dot q_d,A\sp\prime]$.
The Lax equation ${\dot L}=[L,M]$ consequently yields
\begin{eqnarray}
&{{\ddot q}_d\over (a \tau+b \dot q)_d} + {b\over2}\Bigl(
  {{\ddot q}_d\over (a \tau+b \dot q)_d} A +A{{\ddot q}_d\over 
  (a \tau+b \dot q)_d}\Bigr)
  +[{\dot q}_d,A\sp\prime ] \hfill\nonumber\\  &=
[\dot q_d,C]+[A,(B.[a \tau+b \dot q])_d]+A 
(a \tau+b \dot q)_d C -C (a \tau+b \dot q)_d A.\hfill
\label{consistency}
\end{eqnarray}
To solve this equation we
consider the diagonal and off-diagonal terms  separately.
The fact that $A,B,C$ have vanishing diagonal terms results in
\begin{equation}
{{\ddot q}_d\over (a \tau+b \dot q)_d}={\rm Diag}\Big(
    A (a \tau+b \dot q)_d C -C (a \tau+b \dot q)_d A\Big). \label{diag}
\end{equation}
This equation has several consequences. First, observe that if $A$ and $C$
have been determined then we obtain the sought after interaction,
$${\ddot q_j}=\sum_{k\ne j}(a+b \dot q_j) (a+b \dot q_k) V_{jk}(q_j-q_k)=
   (a+b \dot q_j) \Big(A (a \tau+b \dot q)_d C -C (a \tau+b \dot q)_d 
  A\Big)_{jj}.
$$
That is
\begin{equation}
V_{jk}(q_j-q_k)=A_{jk}C_{kj}-C_{jk}A_{kj}=
  \biggl|\matrix{A_{jk}&A_{kj}\cr C_{jk}&C_{kj}\cr}\biggr|
  =-V_{kj}(q_k-q_j).\label{ptla}
\end{equation}
The second and crucial point is that (\ref{diag}) reduces the off-diagonal terms
in our consistency equation (\ref{consistency})\
to a linear equation in $a\tau$ and $b \dot q$.
Utilising our definitions of $G$ and $H$
we find  the off-diagonal terms of (\ref{consistency})\ yield (for $j\ne k$)
$$\biggl(  [\dot q_d,H] + [(G.[a \tau+b \dot q])_d,A]+
   b A (V.[a \tau+b \dot q])_d  +C (a \tau+b \dot q)_d A-A (a \tau+b \dot q)_d C \bigg)_{jk}
   =0.
$$

When $b\ne0$ this equation reduces to considering
\begin{equation}
\biggl(  [\dot q_d,H] +b [(G.\dot q)_d,A]+ b\sp2 A (V.\dot q)_d
   +b C \dot q_d A -b A \dot q_d C\bigg)_{{jk}\atop{j\ne k}}    =0
\label{bnez}
\end{equation}
as the $\tau$ terms of the previous equation
are reproduced by taking $\dot q\propto\tau$.
We will  consider separately the $b=0$ situation which gives the equation
\begin{equation}
\biggl([\dot q_d,H] +a[(B.\tau)_d,A] +a [C,A]\bigg)_{{jk}\atop{j\ne k}}=0.
\label{beqz}
\end{equation}

We may solve the linear equation (\ref{bnez})\  by taking
particular choices for $\dot q$. Substituting $\dot q=e_k$, $\dot q=e_j$ and
$\dot q=e_m$ ($m\ne j,k$) in (\ref{bnez})\ yields the  equations
(\ref{solb}), (\ref{sola}) and (\ref{solc}) respectively.

We have thus shown how the Lax pair determined by the data $A$, $B$ and $C$
fixes the potential $V$ via (\ref{ptla}) and consequently matrices $G$ and
$H$ such that (\ref{solb}), (\ref{sola}) and (\ref{solc}) must hold.
It remains to show that $V$ is also given by (\ref{eq:vjk}). First, if
$A_{jk}=0$ then by (\ref{solb}) and (\ref{eq:defgh}) we find $C_{jk}=0$
and so by (\ref{ptla}) $V_{jk}=0$ as well; thus (\ref{eq:vjk}) holds in this case.
Finally, if $A_{jk}\ne0$, then upon adding (\ref{sola})\ and (\ref{solb})\
we obtain
\begin{equation}
b\,V_{kj}(x)=G_{kj}(x)-G_{jk}(-x).
\label{vodd}
\end{equation}
Making use of the antisymmetry of $V$ together with the definition of $G$
gives as a consequence of (\ref{vodd}) that $B_{kj}(x)=B_{jk}(-x)$.
Equation (\ref{eq:vjk}) now follows from (\ref{ptla}) after making
use of this symmetry and the expression 
\begin{equation}
C_{jk}=A\sp\prime_{jk}- b A_{jk}G_{jk},
\label{defc}
\end{equation}
which follows from (\ref{solb}). Therefore we have established (\ref{eq:vjk}).

The converse of the theorem follows from our initial remarks that the matrices
$A$ and $G$ together with the definition of $V$ suffice to determine the Lax pair.

\end{proof}

Our task therfore is to find matrices $A$ and $G$ for which (\ref{solb}-\ref{eq:vjk})
are satisfied. In order to understand this system of equations it is helpful
to consider the consequences of an entry of $A$ either vanishing identically, or otherwise.
\begin{lem} If $A_{jk}=0$  for some $j,k$ then $H_{jk}=C_{jk}=V_{jk}=0$.
Further, for any $m\ne j,k$ for which $C_{jm}, C_{mk}\ne0$ there exists a
 constant $a_{jmk}$ such that
$ A_{jm}(x)=a_{jmk}C_{jm}(x)$ and $ A_{mk}(y)=a_{jmk}C_{mk}(y) $.
If $a_{jmk}\ne 0$ then
\begin{equation}
A_{jm}(x)=\alpha_{jm}e\sp{x/ a_{jmk}}\,\psi_{jm}(x),\quad\quad
A_{mk}(y)=\alpha_{mk}e\sp{y/ a_{jmk}}\,\psi_{mk}(y).
\label{eq:lema}
\end{equation}
for constants $\alpha_{jm}$, $\alpha_{mk}$.
\end{lem}
\begin{proof}
The vanishing of $H_{jk}$, $C_{jk}$ and $V_{jk}$ is immediate. Further
our assumption means that the right-hand side of (\ref{solc}) vanishes. Because
$C_{jm}$ and $ C_{mk}$ do not vanish identically, (\ref{defc}) entails that neither do
$A_{jm}$ and $A_{mk}$. Thus the first row of the determinant must be proportional
to the second and we have the second assertion of the lemma.
Upon making use of (\ref{defc}) we obtain
$$
A_{jm}(x)=a_{jmk}\big( A\sp\prime_{jm}(x)- b A_{jm}(x)G_{jm}(x)\big)
=a_{jmk}\big( A\sp\prime_{jm}(x)- A_{jm}(x)
{\psi_{jm}\sp\prime(x) \over\psi_{jm}(x)} \big).
$$
When $a_{jmk}\ne0$ this gives
$$
{A\sp\prime_{jm}(x)\over A_{jm}(x)}={1\over a_{jmk}}+
{\psi_{jm}\sp\prime(x) \over\psi_{jm}(x)},
$$
and the first part of (\ref{eq:lema}) follows upon  integration;
the expression for $A_{mk}(y)$ follows similarly.
 
\end{proof}

\begin{lem}
If $A_{jk}\ne0$ for some $j,k$  then
$ V_{kj}(x)=G_{kj}(x)-G_{jk}(-x)$ and $ B_{kj}(x)=B_{jk}(-x)$.
Further, for any $m\ne j,k$:
\begin{enumerate}
\item  if $A_{mk}=0$, then $G_{jm}=G_{km}=c_1$, a constant;
\item  if $A_{mk}\ne0$ and $A_{jm}=0$ then $G_{jm}=G_{mk}=c_2$, a constant; 
\item  if $A_{mk}\ne0$ and $G_{jm}-G_{mk}\ne0$
(and consequently $A_{jm}\ne0$)
then (\ref{solc}) may be written as
\begin{equation}
 A_{jk}={{\biggl|\matrix{A_{jm}&A_{mk}\cr C_{jm}&C_{mk}\cr}\biggr|}\over
           G_{jm}-G_{mk} }
         =b\, A_{jm}A_{mk}+
{{\biggl|\matrix{A_{jm}&A_{mk}\cr A\sp\prime_{jm}&A\sp\prime_{mk}\cr}\biggr|}
\over G_{jm}-G_{mk} }.
\label{sold}
\end{equation}
\end{enumerate}
\end{lem}
\begin{proof}
We have already proven the first part of this lemma in our discussion of the theorem,
(\ref{vodd}). 
Now if $A_{mk}=0$ for some $m$, then $V_{km}=0$ by (\ref{eq:vjk}) and so
only the first term of (\ref{solc}) is nonvanishing. This means $G_{jm}=G_{km}$ and
as these are functions of different arguments, they must be constant.

Assume $A_{mk}\ne0$. Then
employing (\ref{vodd}) for these indices enables us to rewrite (\ref{solc}) as
\begin{equation}
b\big( A_{jk}-b\, A_{jm}A_{mk}\big)\big( G_{jm}-G_{mk}\big)=
b\, 
\biggl|\matrix{A_{jm}&A_{mk}\cr A\sp\prime_{jm}&A\sp\prime_{mk}\cr}\biggr|.
\label{eq:alt}
\end{equation}
The remaining cases are now straightforward.
\end{proof}


Clearly (\ref{sold}) yields a nontrivial relation between the desired functions
$A$ and $G$ and we shall now show that this equation may be recast in the form
(\ref{functional}).
With this aim in mind it is convenient to rewrite (\ref{sold}) in the form
\begin{equation}
\tilde A_{jk}(x+y)=\tilde A_{jm}(x)\tilde A_{mk}(y)-
\psi_{jm}(x)\psi_{mk}(y)
{{\biggl|\matrix{\tilde A_{jm}(x)&\tilde A_{mk}(y)\cr 
         \tilde A\sp\prime_{jm}(x)&\tilde A\sp\prime_{mk}(y)\cr}\biggr|}
\over  
{\biggl|\matrix{ \psi_{jm}(x)& \psi_{mk}(y)\cr 
         \psi\sp\prime_{jm}(x)& \psi\sp\prime_{mk}(y)\cr}\biggr|}},
\label{soldd}
\end{equation}
where $\tilde A=b\, A$. As we have already remarked, when $b\ne0$ we may
rescale so that $b=1$ and for the remainder of this section and until 
section 5 we  assume  that we have done this.

\begin{lem} Equation (\ref{sold}) may be rewritten in the form
(\ref{functional}) where either
\begin{equation}
A_{jk}(x+y)={ { \biggl|\matrix{
  {A_{jm}(x)/ \psi_{jm}(x)}&{A_{jm}(y)/ \psi_{jm}(y)}\cr
  {A_{mk}(x)/ \psi_{mk}(x)}&{A_{mk}(y)/ \psi_{mk}(y)}\cr }\biggr|}
                \over { \biggl|\matrix{
  {1/ \psi_{jm}(x)}&{1/ \psi_{jm}(y)}\cr
  {1/ \psi_{mk}(x)}&{1/ \psi_{mk}(y)}\cr }\biggr|} }
\label{dfd}
\end{equation}
or 
\begin{equation}
A_{jk}(x+y)={ { \biggl|\matrix{
  c_2 A_{mk}(x)&c_2 A_{mk}(y)\cr C_{mk}(x)&C_{mk}(y)\cr }\biggr|}\over
 { \biggl|\matrix{ G_{mk}(x)&G_{mk}(y)\cr 1&1\cr}\biggr|} },
\label{dfe}
\end{equation}
for some constant $c_2$
according to whether
$\biggl|\matrix{\psi_{jm}(x)&\psi_{jm}(y)\cr \psi_{mk}(x)&\psi_{mk}(y)\cr}
\biggr|\ne 0$
vanishes or not.
\end{lem}
{\em Proof}.
Adopting the shorthand $\partial=\partial_j+\partial_k$ observe that
$$\partial\, A_{jm}A_{mk}=-
{\biggl|\matrix{A_{jm}&A_{mk}\cr A\sp\prime_{jm}&A\sp\prime_{mk}\cr}\biggr|}
.$$
Thus 
we may rewrite (\ref{soldd})\ as
\begin{equation}
 A_{jk}=A_{jm}A_{mk}-\psi_{jm}\psi_{mk}
   {\partial\, A_{jm}A_{mk}\over {\partial\, \psi_{jm}\psi_{mk} }}.
\label{dfa}
\end{equation}
Now because $A_{jk}=A_{jk}(x_j-x_k)$ we have $\partial A_{jk}=0$.
Thus applying $\partial$ to both sides of (\ref{dfa})\ shows
$$\partial\, {\partial\, A_{jm}A_{mk}\over {\partial\, \psi_{jm}\psi_{mk} }}
=0,
$$
and consequently the ratio here is a function of $x_j-x_k$, {\it i.e.} 
${\partial\, A_{jm}A_{mk}\over {\partial\, \psi_{jm}\psi_{mk} }}\equiv
A_{jmk}(x_j-x_k)$. If we set $x=x_j-x_m$ and $y=x_m-x_k$ then (\ref{dfa})\
takes the form
\begin{equation} A_{jk}(x+y)=A_{jm}(x)A_{mk}(y)-\psi_{jm}(x)\psi_{mk}(y)A_{jmk}(x+y) .
\label{dfb}
\end{equation}
The left hand side of this equation is symmetric under the interchange of $x$
and $y$. Performing this interchange and subtracting the resulting
equation from (\ref{dfb})\ shows
\begin{equation}
0=\biggl|\matrix{A_{jm}(x)&A_{jm}(y)\cr A_{mk}(x)&A_{mk}(y)\cr}\biggr| -
\biggl|\matrix{\psi_{jm}(x)&\psi_{jm}(y)\cr \psi_{mk}(x)&\psi_{mk}(y)\cr}\biggr|
    A_{jmk}(x+y).
\label{dfc}
\end{equation}

Two possibilities now arise, each leading to an equation of the form 
(\ref{functional}). Suppose first that
$\biggl|\matrix{\psi_{jm}(x)&\psi_{jm}(y)\cr \psi_{mk}(x)&\psi_{mk}(y)\cr}
\biggr|\ne 0$. Then
$$ A_{jmk}(x+y)=
{\biggl|\matrix{A_{jm}(x)&A_{jm}(y)\cr A_{mk}(x)&A_{mk}(y)\cr}\biggr| \over
\biggl|\matrix{\psi_{jm}(x)&\psi_{jm}(y)\cr \psi_{mk}(x)&\psi_{mk}(y)\cr}
\biggr|}.
$$
Upon substituting this expression into (\ref{dfb})\ one obtains
(\ref{dfd}).

In the case when 
$\biggl|\matrix{\psi_{jm}(x)&\psi_{jm}(y)\cr \psi_{mk}(x)&\psi_{mk}(y)\cr}
\biggr|=0$ we have $\psi_{jm}(x)=c_1 \psi_{mk}(x)$ for some constant
$c_1$.  Therefore $G_{jm}(x)=G_{mk}(x)$.
Likewise from (\ref{dfc})\  we have that $A_{jm}(x)=c_2 A_{mk}(x)$ and
hence $C_{jm}(x)=c_2 C_{mk}(x)$.
Substituting these relations into (\ref{sold})\ obtains (\ref{dfe}),
again of the stated form.

{\hfill $\square$}

\section{The Functional Equation}
Thus far we have reduced the consistency requirements for the Lax pair
(\ref{ansatz})\ to a functional equation of the form 
(\ref{functional}) and this section looks briefly at this equation.
Particular cases of this equation have been described in the 
literature \cite{BCb,OPc}.
Here we describe the  invariances ${\cal G}$  of (\ref{functional}).
Theorem $1$ gives a representative of each  ${\cal G}$  orbit
on the solutions of (\ref{functional})\ with a particularly nice form.
For later calculations it is instructive to see how the
stated solution satisfies (\ref{functional}), an exercise involving
some elliptic function identities.
We end the section by deriving the relativistic Calogero-Moser model found by
Bruschi and Calogero \cite{BCa}.

First observe that a large group of symmetries ${\cal G}$ act on the solutions
of (\ref{functional}). The transformation
\begin{equation}
\Biggl( \phi_1(x), {\phi_2(x)\choose\phi_3(x)},{\phi_4(x)\choose\phi_5(x)} 
\Biggr) \rightarrow
  \Biggl( c\, e\sp{\lambda x} \phi_1(x), U {e\sp{-\lambda\sp\prime x}
\phi_2(x)\choose e\sp{-\lambda\sp\prime x}\phi_3(x)},
V {e\sp{\lambda\sp{\prime\prime}x} \phi_4(x)\choose
   e\sp{\lambda\sp{\prime\prime}x} \phi_5(x)} \Biggr)
\label{symm}
\end{equation}
clearly preserves (\ref{functional})\ provided
\begin{equation}
\lambda+\lambda\sp\prime+\lambda\sp{\prime\prime}=0,\quad\quad
  U,V\in GL_2,\quad\quad {\rm and}\quad\quad \det U=c\,\det V.
\label{constraints}
\end{equation}
Further, (\ref{functional}) is also preserved by
\begin{equation}
\Biggl( \phi_1(x), {\phi_2(x)\choose\phi_3(x)},{\phi_4(x)\choose\phi_5(x)}
\Biggr) \rightarrow\Biggl(
{1\over \phi_1(x)},{\phi_4(x)\choose\phi_5(x)},{\phi_2(x)\choose\phi_3(x)}
\Biggr)
\label{symms}
\end{equation}
and
\begin{equation}
\Biggl( \phi_1(x), {\phi_2(x)\choose\phi_3(x)},{\phi_4(x)\choose\phi_5(x)}
\Biggr) \rightarrow\Biggl(\phi_1(x),
f(x){\phi_2(x)\choose\phi_3(x)},f(x){\phi_4(x)\choose\phi_5(x)}\Biggr).
\label{symmsb}
\end{equation}

These symmetries enable one to find a
solution of (\ref{functional})\ on each  ${\cal G}$  orbit
with a particularly nice form.
It is instructive
to see how the stated solution satisfies (\ref{functional}).
>From the definition of the zeta function we have
$$ \bigr(\ln \Phi(x;\nu)\bigl)\sp\prime = -\zeta(\nu-x)-\zeta(x)+\zeta(\nu).
$$
Thus
\begin{eqnarray*}
 {\biggl| \matrix{\Phi(x;\nu)&\Phi(y;\nu)\cr\Phi(x;\nu)\sp\prime&
     \Phi(y;\nu)\sp\prime\cr}\biggr| }
&=&{\Phi(x;\nu)\Phi(y;\nu)\biggl[ \bigr(\ln \Phi(y;\nu)\bigl)\sp\prime
  -\bigr(\ln \Phi(x;\nu)\bigl)\sp\prime\biggr] } \hfill\\
&=&\Phi(x;\nu)\Phi(y;\nu)\biggl[
\zeta(\nu-x)+\zeta(x)+\zeta(-y)+\zeta(y-\nu)\biggr]. \hfill
\end{eqnarray*}
Upon using the definition of $\Phi$ the right hand side of this
equation takes the form
\begin{equation}
\Phi(x+y;\nu){\sigma(\nu-x)\sigma(\nu-y)\sigma(x+y)\over
    \sigma(\nu-x-y)\sigma(\nu)\sigma(x)\sigma(y) }
    \biggl[ \zeta(\nu-x)+\zeta(x)+\zeta(-y)+\zeta(y-\nu)\biggr].
\label{rhs}
\end{equation}
After noting  the two identities \cite{WW}
\begin{equation}
\zeta(x)+\zeta(y)+\zeta(z)-\zeta(x+y+z)=
{\sigma(x+y)\sigma(y+z)\sigma(z+x)\over\sigma(x)\sigma(y)\sigma(z)\sigma(x+y+z)}
\label{eq:zetas}
\end{equation}
and 
\begin{equation}
\wp (x)-\wp(y)=
{\sigma(y-x)\sigma(y+x)\over\sigma\sp2(y)\sigma\sp2(x)}
\label{eq:wps}
\end{equation}
we find (\ref{rhs})  simplifies to
$\Phi(x+y;\nu)\bigl[{\wp }(x)-{\wp }(y)\bigr] $,
where ${\wp }(x)=-\zeta\sp\prime(x)$ is the Weierstrass ${\wp }$-function. 
Putting these together yields the addition formula
\begin{equation}
\Phi(x+y;\nu)=
{{\biggl| \matrix{\Phi(x;\nu)&\Phi(y;\nu)\cr\Phi(x;\nu)\sp\prime&
     \Phi(y;\nu)\sp\prime\cr}\biggr| }\over
{\wp }(x)-{\wp }(y) },
\label{addn}
\end{equation}
and consequently a solution of (\ref{functional})\ with the stated form. 

A consequence of Theorem $1$ then is 
\begin{col}
If $A_{jk}$ satisfies (\ref{sold}) then
\begin{equation}
A_{jk}(x)=c_{jk}{\Phi(x,\nu_{jk})\over\Phi(x,\mu_{jk})} e\sp{\lambda_{jk}x}
\label{eq:solform}
\end{equation}
for some constants $c_{jk}$, $\nu_{jk}$, $\mu_{jk}$ and $\lambda_{jk}$.
\end{col}

\noindent{\bf Example} The relativistic
example of Calogero and
Bruschi arises as a particular case of our ansatz
when $a=0,\ b=1$ and 
$$A_{jk}(x)=(1-\delta_{jk})\alpha(x),\quad
  B_{jk}(x)=(1-\delta_{jk})\beta(x),\quad
  C_{jk}(x)=(1-\delta_{jk})\gamma(x).
$$
In this case (\ref{dfe})\ takes the form
\begin{equation}
\alpha(x+y)=
{ { \biggl|\matrix{
   \alpha(x)& \alpha(y)\cr \gamma(x)&\gamma(y)\cr }\biggr|}\over
 { \biggl|\matrix{ \eta(x)&\eta(y)\cr 1&1\cr}\biggr|} },
\label{exa}
\end{equation}
where $\eta(x)=\beta(x)+{1\over2}v(x)$. Comparison of (\ref{exa})\ 
with the general solution
of (\ref{functional})\ shows the solution to be given by
$$\alpha(x)={\Phi(x;\nu)\over\Phi(x;\mu)}\quad\quad
  \eta(x)=-\bigr(\ln \Phi(x;\mu)\bigl)\sp\prime=
  -{1\over2}{{\wp\prime(x)-\wp\prime(\mu)}\over{\wp(x)-\wp(\mu)}}.
$$
In this case $\alpha(x)\alpha(-x)=\big(\wp(x)-\wp(\nu)\big)/
\big(\wp(x)-\wp(\mu)\big)$ and so by (\ref{eq:vjk}) 
$$v(x)={{\wp\prime(x)}\over{\wp(\mu)-\wp(x)}},$$
and we have recovered\footnote{
We remark in passing that the the scaling of the elliptic function
$$
\wp(t\, x| t\,\omega,t\,\omega\sp\prime)=t\sp{-2}\wp(x|\omega,\omega\sp\prime)
$$
is accounted for by the scaling of the Lax operator
(\ref{eq:Laxsymm}) and doesnt appear to give any new potentials.} the
results of  references  \cite{Ra} and  \cite{BCa}.
 
\section{Application of the Functional Equation}

Thus far we have discussed the functional equation (\ref{functional}) in 
isolation while our application involves a system of such equations.
In solving this system we  encounter further constraints. To see
how these arise consider (\ref{dfd}). Our theorem says
$$
A_{jk}(x)=c_{jk}{\Phi(x;\nu_{jk})\over\Phi(x;\mu_{jk})} e\sp{\lambda_{jk}x}
$$
(for some constants $c_{jk}$, $\nu_{jk}$, $\mu_{jk}$ and $\lambda_{jk}$)
and further that
$$
\pmatrix{ {A_{jm}(x)/ \psi_{jm}(x)}\cr {A_{mk}(x)/ \psi_{mk}(x)} \cr}
=f(x)e\sp{-\lambda\sp\prime x} U 
\pmatrix{ \Phi(x,\nu_{jk})\cr \Phi\sp\prime(x,\nu_{jk})\cr},
$$
$$
\pmatrix{{1/ \psi_{jm}(x)}\cr{1/ \psi_{mk}(x)}\cr}
=f(x)e\sp{\lambda\sp{\prime\prime}x} V
\pmatrix{ \Phi(x,\mu_{jk})\cr \Phi\sp\prime(x,\mu_{jk})\cr}
$$
for an appropriate function $f(x)$ and matrices $U,V$ such that
$$
\lambda_{jk}+\lambda\sp\prime+\lambda\sp{\prime\prime}=0\quad\quad
\det{U}=c_{jk}\det{V}.
$$
Now if $A_{jm}$ and $A_{mk}$ are also given by (\ref{eq:solform})
we encounter restrictions on the possible parameters appearing:
\begin{thm} Let $A_{jk}$, $A_{jm}$ and $A_{mk}$ have the form 
(\ref{eq:solform}) and be related by (\ref{sold}). Then
the constants determining these solutions are related by
\begin{equation}
\nu_{jk}-\mu_{jk}=\nu_{jm}-\mu_{jm}=\nu_{mk}-\mu_{mk}
\label{eq:relmns}
\end{equation}
\begin{equation}
\lambda_{jk}+\zeta(\nu_{jk})-\zeta(\mu_{jk})=
\lambda_{jm}+\zeta(\nu_{jm})-\zeta(\mu_{jm})=
\lambda_{mk}+\zeta(\nu_{mk})-\zeta(\mu_{mk})
\label{eq:rellmns}
\end{equation}
and 
$c_{jk}={\sigma(\nu_{jk})\over \sigma(\mu_{jk})}\tau_{jk}$,
$c_{jm}={\sigma(\nu_{jm})\over \sigma(\mu_{jm})}\tau_{jm}$,
$c_{mk}={\sigma(\nu_{mk})\over \sigma(\mu_{mk})}\tau_{mk}$
where the $\tau$'s satisfy
\begin{equation}
{\tau_{jk}\over\tau_{jm}\tau_{mk}}=
{\sigma(\nu_{jm}+\nu_{mk}-\nu_{jk})\over \sigma(\mu_{jm}+\mu_{mk}-\mu_{jk})}.
\label{eq:relts}
\end{equation}
Letting $\mu_{jk}-\nu_{jk}=c$ and 
$\lambda_{jk}+\zeta(\nu_{jk})-\zeta(\mu_{jk})=\rho$
then
\begin{equation}
A_{jk}=\tau_{jk}{\sigma(\nu_{jk}-x)\over \sigma(c+\nu_{jk}-x)}e\sp{\rho x}
\label{eq:reltas}
\end{equation}
and similarly
$$
A_{jm}=\tau_{jm}{\sigma(\nu_{jm}-x)\over \sigma(c+\nu_{jm}-x)}e\sp{\rho x},
\quad\quad
A_{mk}=\tau_{mk}{\sigma(\nu_{mk}-x)\over \sigma(c+\nu_{mk}-x)}e\sp{\rho x}.
$$
Finally
\begin{equation}
 G_{jm}(x)= -\zeta(x-\mu_{jm})+\zeta(x+\mu_{mk}-\mu_{jk}) +const
\label{eq:finrels}
\end{equation}
and similarly for $G_{mk}(y)$ with the same constant appearing.
\end{thm}

The proof of this theorem is rather lengthy, making repeated use of the
elliptic function identies introduced in the previous section; it is
given in appendix A.  The first two relations (\ref{eq:relmns}) and (\ref{eq:rellmns})
follow by equating  poles and zeros amongst the various terms
while the relation  (\ref{eq:relts}) comes from the determinantal constraint.
The final constraint (\ref{eq:finrels}) arises by considering (\ref{eq:alt}),
which may be recast as
$$
G_{jm}(x)-G_{mk}(y)=
\partial \ln\left(1 -\frac{A_{jm}(x) A_{mk}(y)}{A_{jk}(x+y)}\right).
$$
We remark that when $A_{jm}(x)=c_2 A_{mk}(x)$ (and so
$\nu_{jm}=\nu_{mk}$, $\mu_{jm}=\mu_{mk}$, $\lambda_{jm}=\lambda_{mk}$,
$c_{jm}=c_2 c_{mk}$) several of these relations are immediately satisfied.

It is worth reflecting on this theorem. Given any three $A_{jk}$, 
$A_{jm}$ and $A_{mk}$  of the form (\ref{eq:solform})
and connected via (\ref{sold}), the constants determining these functions are
restricted.
In particular, suppose {\em every} entry of the Lax matrix $A$ is nonvanishing
and $G_{jm}-G_{mk}\ne0$ for every triple of distinct indices. Then the theorem
holds for every triple $A_{jk}$, $A_{jm}$ and $A_{mk}$. Consideration of
(\ref{eq:finrels}) shows that if this is to define a function $G_{jm}$
then
\begin{equation}
\mu_{jm}+\mu_{mk}-\mu_{jk}=\mu
\label{eq:homo}
\end{equation}
for some fixed constant $\mu$ and every distinct triple $j,m,k$.
Now (\ref{eq:homo}) holds for every distinct triple if and only if
(for each $j,k$)
$$
\mu_{jk}=\mu+\mu_j -\mu_k
$$
and similarly  $\nu_{jk}=\nu+\nu_j -\nu_k$. In this case
$\tau_{jk}=\sigma(\mu)/\sigma(\nu)$ and so
$$
A_{jk}(x)={\sigma(\mu)\over\sigma(\nu)}
{\sigma(\nu+\nu_j -\nu_k-x)\over \sigma(\mu+\nu_j -\nu_k-x)}e\sp{\rho x}.
$$
That is
$$
A_{jk}(x_j-x_k)={\Phi(x_j-\nu_j-(x_k-\nu_k);\nu)\over 
\Phi(x_j-\nu_j-(x_k-\nu_k);\mu)}e\sp{[\rho-\zeta(\nu)+\zeta(\mu)] x}
$$
and we have recovered the relativistic Calogero-Moser model described
in the last section. Now $G_{jm}=G_{mk}\Leftrightarrow G_{jm}=G_{mk}=constant$.
Thus we obtain the following \lq generic\rq\ description of the 
relativistic Calogero-Moser model:

\begin{thm}
A Lax Pair of the form (\ref{ansatz}) for which the matrix $A$ has no
vanishing entries and for which the matrix $G$ has nonconstant entries
describes the relativistic Calogero-Moser model.
\end{thm}

We have just considered the situation where  every entry of $A$ satisfying (\ref{dfd}).
We conclude the section by considering the opposite extreme where no entry does.

\noindent{\bf Example}
Here we adopt the ansatz
$$
A=\pmatrix{0&a_1&0&0&\ldots&0&0\cr
           1&0&a_2&0&      &0&0\cr
           1&1&0&a_3&      &0&0\cr
       \vdots&\cr
           1&1&1&1&        &0&a_{n-1}\cr
      1&1&1&1&             &1&0\cr}
$$
and where we shall determine the $a_i'{\rm s}\ne0$.
We will work through the various cases determined by Lemma 1.
\begin{enumerate}
\item
$j<k-1$. Then $A_{jk}=0$ and the only possible nonzero term in
(\ref{solc}) is
$$
0=\biggl|\matrix{A_{jj+1}&A_{j+1\, j+2}\cr C_{jj+1}&C_{j+1\, j+2}\cr}\biggr|.
$$
Thus for each $j< n-1$ 
$$
{C_{jj+1}\over A_{jj+1}}={C_{j+1\, j+2}\over A_{j+1\, j+2}}=\lambda
$$
and consequently for each $j<n$
\begin{equation}
{ A\sp\prime_{jj+1}\over  A_{jj+1}}-G_{jj+1}=\lambda.
\label{rt1}
\end{equation}
\item
$j=k-1$. In this case $A_{jj+1}\ne0$  and now  if
\begin{enumerate}
\item
$m<j$ then $A_{mk}=0$ and $G_{jm}=G_{j+1\, m}$.
\item
$j+2<m$ then $G_{jm}=G_{j+1\, m}$.
\item $m=j+2$ then $G_{jj+2}-G_{j+1\, j+2}+V_{j+1\, j+2}=0$ whence upon
using (\ref{eq:vjk})
\begin{equation}
G_{jj+2}-G_{j+1\, j+2}={A\sp\prime_{j+1\, j+2}\over 1-A_{j+1\, j+2}}.
\label{rt2}
\end{equation}
\end{enumerate}
\item
$k<j$. Then $A_{jk}=1$. Now (\ref{eq:alt}) becomes
$$
\big( 1- A_{jm}A_{mk}\big)\big( G_{jm}-G_{mk}\big)=
\biggl|\matrix{A_{jm}&A_{mk}\cr A\sp\prime_{jm}&A\sp\prime_{mk}\cr}\biggr|.
$$
and we find
\begin{enumerate}
\item
if $m< k-1$ or $k<j<m-1$ that $G_{jm}=G_{mk}$,
\item
$m=k-1$
\begin{equation}
G_{jk-1}-G_{k-1\, k}={A\sp\prime_{k-1\, k}\over 1-A_{k-1\, k}}
\label{rt3}
\end{equation}
\item
$j=m-1$
\begin{equation}
G_{jj+1}-G_{j+1\, k}=-{A\sp\prime_{j j+1}\over 1-A_{j j+1}}
\label{rt4}
\end{equation}
\end{enumerate}
with no constraints arising when $k<m<j$.
Further, when $k<j+1$, we have $V_{jk}=0=G_{jk}-G_{kj}$
and so
\begin{equation}
G_{jk}=G_{kj}\quad\quad |j-k|>1.
\label{rt5}
\end{equation}
\end{enumerate}

Now case (2a) tells us each column of the matrix $G$ is constant
below the diagonal while (2b) tells us each column above the
superdiagonal is constant. Combining this information with 
(\ref{rt5}) enables us to parameterize $G$ as
$$
G=\pmatrix{0&g_1&0&0&\ldots&0&0\cr
           1&0&g_2&0&      &0&0\cr
           1&1&0&g_3&      &0&0\cr
       \vdots&\cr
           1&1&1&1&        &0&g_{n-1}\cr
      1&1&1&1&             &1&0\cr}
+d\ 
\pmatrix{0&1&1&1&\ldots&1&1\cr
           1&0&1&1&      &1&1\cr
           1&1&0&1&      &1&1\cr
       \vdots&\cr
           1&1&1&1&        &0&1\cr
      1&1&1&1&             &1&0\cr}
$$
where $d$ is a constant. Comparison with (\ref{rt1})-(\ref{rt4})
shows we are left with two equations,
$$
g_j={-a\sp\prime_j \over{1-a_j}}\equiv V_{jj+1}\quad\quad
{a\sp\prime_j\over a_j}-g_j=\lambda+d,
$$
with solution
$$
a_j(x)=1-{1\over 1+c_{jj+1}e\sp{(\lambda+d)x} }
$$
and $g_j(x)=-(\lambda+d)a_j(x)$. Here $c_{jj+1}$ is a constant that
may be removed by shifting the $x$'s.

We have recovered the relativistic Toda lattice of \cite{Rb}.
Our construction has given the Lax pair ($b=1$, $a=0$)
$L=\sqrt{\dot q_d}\,(I+A)\sqrt{\dot q_d}$ and
$$
M=(B.\dot q)_d+\sqrt{\dot q_d}\, C\sqrt{\dot q_d}
=-d\, L+ \big(d\sum_{j=1}\sp{n}\dot q\big)I
+
\pmatrix{ {1\over2}g_1 \dot q_2&0&&0\cr
          0& {1\over2}g_1 \dot q_1+{1\over2}g_2 \dot q_3&\cr
          0&&\ddots&\cr}
$$
$$
+\sqrt{\dot q_d}
\pmatrix{0&-g_1&0&\ldots&0&0\cr
           0&0&-g_2&      &0&0\cr
       \vdots&\cr
           0&0&0&        &0&-g_{n-1}\cr
      0&0&0&             &0&0\cr}
\sqrt{\dot q_d} .
$$
Upon defining a conjugate Lax pair $L_N$, $M_N$ by
$L=\sqrt{\dot q_d} \, L_N\, 1/\sqrt{\dot q_d}$ and
$M_N=1/ \sqrt{\dot q_d}\big(M+d\, L-\big(d\sum_{j=1}\sp{n}\dot q\big)I
\big) \sqrt{\dot q_d}+\dot{\sqrt{\dot q_d}} / \sqrt{\dot q_d}$
we obtain the Lax pair
$$
L_N=
\pmatrix{\dot q_1&a_1\dot q_2&0&\ldots&&0\cr
           \dot q_1&\dot q_2&a_2\dot q_3&      &&0\cr
           \dot q_1&\dot q_2&\dot q_3&&      &&0\cr
       \vdots&\cr
           \dot q_1&\dot q_2&\dot q_3&&        &&a_{n-1}\dot q_n\cr
      \dot q_1&\dot q_2&\dot q_3&&             &&\dot q_n\cr},
$$
$$
M_N=
\pmatrix{g_1\dot q_2&-g_1\dot q_2&0&\ldots&0\cr
           0&g_2\dot q_3&-g_2\dot q_3&     &0\cr
       \vdots&\cr
           0&0&0&        &-g_{n-1}\dot q_n\cr
      0&0&0&             &0\cr}
$$
When $\lambda=1$ and $d=0$ this is the Lax pair of \cite{BRa}.

\section{The Nonrelativistic case.}
Our discussion has thus far  focussed on the $b\ne0$ case of our
ansatz for the Lax-pair and this has yielded the relativistic
Calogero-Moser and Toda systems. We shall now consider the case $b=0$
and see that this includes the nonrelativistic limits of these systems.
The nonrelativistic dynamics is less constrained than the relativistic
situation and this reflects itself in a more complicated functional
equation which we have not been able to solve in general. In this
section we shall first obtain the relevant equation and show how
it encompasses the nonrelativistic Calogero-Moser and Toda systems as
well as that of Buchstaber and Perelomov \cite{BP}. Our approach
leads to a new derivation of the Calogero-Moser model in which we
also determine the various symmetry properties of the functions
entering the ansatz.
Recalling that the diagonal entries of $A,\, B\in {\cal M}_n$ vanish
we begin with:

\begin{thm}
Equation (\ref{beqz})\ yields the functional equation
\begin{equation}
\sum_{l}A_{jk}(B_{jl}-B_{kl}) + A\sp\prime_{jl}A_{lk}
  -A_{jl}A\sp\prime_{lk}=0.
\label{solg}
\end{equation}
Further, each of the functions
$\Psi_{jlk}\equiv A_{jk}(B_{jl}-B_{kl}) + A\sp\prime_{jl}A_{lk}
  -A_{jl}A\sp\prime_{lk}$
appearing as the terms of this sum, depend only on the combination
$q_j-q_k$. Thus with $x=q_j-q_l$ and $y=q_l-q_k$ we have
\begin{equation}
\Psi_{jlk}(x+y)=A_{jk}(x+y)\big( B_{jl}(x)-B_{kl}(-y)\big) +
A\sp\prime_{jl}(x)A_{lk}(y)-A_{jl}(x)A\sp\prime_{lk}(y).
\label{eqsolq}
\end{equation}
\end{thm}
\begin{proof}
In order for (\ref{beqz})\ to remain true for all $\dot q$ we must have
\begin{equation}
H_{jk}=0\label{sole}
\end{equation}
and therefore $C_{jk}=A\sp\prime_{jk}$. In fact this is the
$b\rightarrow0$ limit of equations (\ref{solb}) and (\ref{sola}).
With these simplifications (\ref{beqz}) then becomes
$$
A_{jk}(B_{jk}-B_{kj})+
   \sum_{l\atop{l\ne j,k}}A_{jk}(B_{jl}-B_{kl}) + A\sp\prime_{jl}A_{lk}
  -A_{jl}A\sp\prime_{lk}=0,
$$
and with the conventions stated above we have (\ref{solg}).

Now the only dependence on $q_l$ ($l\ne j,k$) in (\ref{solg})
comes from the $l$-th term of the sum. Thus upon
taking the derivative $\partial_l$ of this equation we see that
$$
0=\partial_l\big( A_{jk}(B_{jl}-B_{kl}) + A\sp\prime_{jl}A_{lk}
  -A_{jl}A\sp\prime_{lk}\big)=
 -(\partial_j+\partial_k)\big( A_{jk}(B_{jl}-B_{kl}) + A\sp\prime_{jl}A_{lk}
  -A_{jl}A\sp\prime_{lk}\big)
$$
from which we may conclude that $\Psi_{jlk}=\Psi_{jlk}(q_j-q_k)$
as stated. We remark that  the $b\rightarrow0$ limit of equation (\ref{solc}) divided
by $b$ gives the quantity $\Psi_{jmk}$.
\end{proof}
\begin{col}
Solutions of (\ref{eqsolq}) satisfy the functional equation (\ref{Bfun}) with $N=2$.
\end{col}
\begin{proof}
Upon interchanging $x$ and $y$ in (\ref{eqsolq}) and subtracting we obtain
\begin{eqnarray}
A_{jk}(x+y) &{ \biggl|\matrix{
 B_{jl}(x)+B_{kl}(-x)&B_{jl}(y)+B_{kl}(-y)\cr 1&1\cr}\biggr|}
+{ \biggl|\matrix{A\sp\prime_{jl}(x)&A\sp\prime_{jl}(y)\cr
                  A_{lk}(x)&A_{lk}(y)\cr}\biggr|}\\
&+{ \biggl|\matrix{A\sp\prime_{lk}(x)&A\sp\prime_{lk}(y)\cr
                  A_{jl}(x)&A_{jl}(y)\cr}\biggr|}
=0\hfill\\
\label{functional3}
\end{eqnarray}
Simple rearrangement of this gives equations of the form
(\ref{functional3}) or of
(\ref{Bfun}) with $N=2$.

\end{proof}

Before further investigating (\ref{solg}) it is instructive to
see how the nonrelativistic Calogero-Moser and Toda systems arise
in this context and present a new example.
\vskip0.2in
\noindent{\bf Example}
The Calogero-Moser reduction follows when we assume the functions
$A_{jk}$ and $B_{jk}$ dont depend on the indices $j,k$. Upon setting
$A_{jk}(q_j-q_k)=A(q_j-q_k)$ and $B_{jk}(q_j-q_k)=B(q_j-q_k)$
the function $\Psi_{jlk}$ defined above takes the form
$$
\Psi_{jlk}\equiv A(q_j-q_k)\big(B(q_j-q_l)-B(q_k-q_l)\big) +
  A\sp\prime(q_j-q_l)A(q_l-q_k)-A(q_j-q_l)A\sp\prime(q_l-q_k).
$$
Now the lemma asserts that $\Psi_{jlk}$ is independent of $q_l$
($l\ne j,k$) and so (for example by setting $q_l=0=q_{l\sp\prime}$)
we see that $\Psi_{jlk}=\Psi_{j{l\sp\prime}k}$ for $l,{l\sp\prime}\ne j,k$.
Thus (\ref{solg}) takes the form
\begin{eqnarray}
\nonumber
0&=&A(x+y)\big({B(x+y)-B(-x-y)}\big) +\\ 
&&\quad
(n-2)\big( A(x+y)\big[ B(x)-B(-y)\big] +A\sp\prime(x)A(y)-A(x)A\sp\prime(y)
\big),
\label{eqcalmo}
\end{eqnarray}
where $x=q_j$ and $y=-q_k$.
Letting $B_o$ denote the odd part of $B$, then upon interchanging
$x$ and $y$ in (\ref{eqcalmo}) and adding leads to 
$$
0=A(x+y)\big( B_o(x+y)+ {(n-2)\over2}\big[ B_o(x)+B_o(y)\big]\big).
$$
Now the functional equation ($n>1$)
$$
0=B_o(x+y)+ {(n-2)\over2}\big[ B_o(x)+B_o(y)\big]
$$
only has $B_o=0$ as a solution and so if $A(x)\not\equiv0$ we
may assume $B$ in (\ref{eqcalmo}) is an even function. In this
case the leading term vanishes and we are left with
$$
0=A(x+y)\big[ B(x)-B(y)\big] +A\sp\prime(x)A(y)-A(x)A\sp\prime(y)
$$
which is the functional equation (\ref{calfun}) obtained by
Calogero \cite{Ca2} and that has been variously solved
\cite{Ca2, OPc, PS, BCb, Bu1}. Observe that by setting $y=-x$ in this 
equation and using the fact that $B(x)$ is an even function we may
deduce that $A(x)$ is an odd function and so we have obtained the
symmetries of $A$ and $B$ normally
\cite{Ca2, OPc} imposed as constraints on the Lax-pair.

\noindent{\bf Example} The Toda reduction follows by assuming
$B_{jk}$ and $\Psi_{jlk}$ vanish for all possible distinct
indices. From this we deduce that
$$ A_{jk}(x)=\alpha_{jk}\, e\sp{\lambda_{jk}x}$$
for some constants $\alpha_{jk}$ and $\lambda_{jk}$. The vanishing of
$\Psi_{jlk}$ relates these constants (for $j\ne l\ne k$) by
\begin{equation}
(\lambda_{jl}-\lambda_{lk})\alpha_{jl}\alpha_{lk}=0
\label{eqtodac1}
\end{equation}
while the nonvanishing of an interaction $V_{jk}$ means (using (\ref{ptla}))
that
\begin{equation}
(\lambda_{jk}-\lambda_{kj})\alpha_{jk}\alpha_{kj}\ne0.
\label{eqtodac2}
\end{equation}
We thus seek solutions of (\ref{eqtodac1}) and (\ref{eqtodac2}).
Observe that if $V_{jl}$ and $V_{lk}$ are both nonvanishing
then $\lambda_{jl}=\lambda_{lk}\ne \lambda_{kl}=\lambda_{lj}$. 
Consequently if $V_{ij}$, $V_{jl}$ and $V_{lk}$ are nonvanishing
then (for example) $V_{jk}=0$. To see this suppose to the contrary and note
that $\Psi_{ijk}=0$ means $\lambda_{ij}=\lambda_{jk}$, while
$\Psi_{jkl}=0$ means $\lambda_{jk}=\lambda_{kl}$
and so $\lambda_{ij}=\lambda_{kl}$. However our previous observation
shows $\lambda_{ij}=\ldots=\lambda_{lk}\ne\lambda_{kl}$, yielding
a contradiction. Therefore $V_{jk}=0$. Our ansatz means we can only
achieve (possibly cyclic) chains of nonvanishing interactions,
$V_{12}=V_{23}=\ldots=V_{n-1\,n}$ ($=V_{n1}$). Setting
$\alpha_{i\, i+1}=1=\alpha_{i+1\, i}$ and 
$\lambda_{i\, i+1}=-\lambda_{i+1\, i}$ say, with all other $\alpha_{ij}$
and $\lambda_{ij}$ vanishing, yields the (periodic) Toda lattice.
Equally the work of \cite{Ino} shows how to obtain the Toda
systems as limits of Calogero-Moser models.
\noindent{\bf Example} By taking 
\begin{equation}
A_{jk}(x)=\Phi(x;\nu)\, a_k,\quad\quad B_{jk}(x)=\wp(x)\, a_k
\label{newwy}
\end{equation}
where the $a_k$ are constants, we find
$$
\Psi_{jlk}(x+y)=\Phi(x+y;\nu)\left( \wp(x)-\wp(y)\right)\, a_k\, a_l -
{{\biggl| \matrix{\Phi(x;\nu)&\Phi(y;\nu)\cr\Phi(x;\nu)\sp\prime&
     \Phi(y;\nu)\sp\prime\cr}\biggr| }}\, a_k\, a_l =0.
$$
Here we have used (\ref{addn}) to show the vanishing of $\Psi_{jlk}$.
By Theorem $5$ we therfore have a new Lax pair associated with the
potentials (\ref{eq:newnonrel})
$$
V_{jk}(x)=a_j a_k\,  \wp\sp\prime(x).
$$
When $a_j=a_k$ this yields the usual type IV
Calogero-Moser potential.

\vskip0.2in

Although we cannot as yet solve (\ref{solg}) or (\ref{eqsolq}) in general,
we can say a little more according to whether $A_{jk}(B_{jl}-B_{kl})=0$
or $A_{jk}\ne0$. For the first we note:

\begin{lem}
The functional equation
\begin{equation}
F(x+y)=\phi(x)\psi\sp\prime(y)-\phi\sp\prime(x)\psi(y)
\label{eqlm4f}
\end{equation}
has,  up to symmetries, the solution $F(x)=c_1 c_3\exp(\lambda x)$, 
$\phi(x)=c_1\exp(\lambda x)$, $\psi(x)=(c_2+c_3 x)\exp(\lambda x)$.
\end{lem}
This equation is  a particular case of the equation
$$
f(x+y)+g(x-y)=\sum_{j=1}\sp{n} f_j(x)g_j(y)
$$
which has a long history \cite{Acz}; the solution, along standard lines,
is given in Appendix B.
In our context it yields
\begin{col}
If $A_{jk}(B_{jl}-B_{kl})=0$ the solution  to (\ref{eqsolq}) takes the
form $\Psi_{jlk}(x)=(c_1 c_3 -c_2 c_4)\exp(\lambda x)$,
$A_{jl}(x)=(c_2+c_3 x)\exp(\lambda x)$ and
$A_{lk}(y)=(c_1+c_4 y)\exp(\lambda y)$, where  $c_3 c_4=0$.
\end{col}
\noindent{\it Remark:} The Toda reduction was a particular example of
this corollary. The various constants appearing are not all independent, but
related by (\ref{solg}).
\vskip0.2in

We now suppose  that $A_{jk}\ne0$. Upon setting
$$
\Phi_{jk}= \sum_m 
{A\sp\prime_{jm}A_{mk} -A_{jm}A\sp\prime_{mk}	\over A_{jk}}
$$
for $j\ne k$ and $\Phi_{jj}=0$ we may rewrite (\ref{solg}) as
\begin{equation}
\sum_m\big( B_{jm}-B_{km}\big) +\Phi_{jk}=0.
\label{eq:phis}
\end{equation}
\begin{lem}
For each $j$, $k$ and $l$ for which $A_{jk}, A_{kl}$ and
$A_{lj}$ are nonvanishing then (\ref{solg}) yields the functional
equation
\begin{equation}
\Phi_{jk}+\Phi_{kl}+\Phi_{lj}=0
\label{eq:phis3}
\end{equation}
and consequently
\begin{equation}
\Phi_{jk}+\Phi_{kj}=0
\label{eq:phisanti}
\end{equation}
\end{lem}
{\em Proof}.
This follows from (\ref{eq:phis}) as
$$
B_{jm}-B_{km}+B_{km}-B_{lm}+B_{lm}-B_{jm}=0.
$$
{\hfill $\square$}

\vskip0.2in
\noindent{\bf Example}
In the case when $n=3$, equation  (\ref{eq:phis3})
reduces to the functional equation of Buchstaber and Perelomov \cite{BP},
\begin{equation}
\big(\, f(x)+g(y)+h(z)\,\big)\sp2=F(x)+G(y)+H(z)\quad\quad
x+y+z=0.
\end{equation}
The equation is related to the factorization of a three-body
quantum mechanical ground-state wavefunction \cite{Su1,Su2,Ca3}.
In this case  there is a unique $m\ne j,k$ such that
$$\Phi_{jk}=
{A\sp\prime_{jm}A_{mk} -A_{jm}A\sp\prime_{mk}   \over A_{jk}}.
$$
Set $x=q_2-q_3$, $y=q_3-q_1$, $z=q_1-q_2$ (and so $x+y+z=0$)
and
$$
f(x)=-A_{23}(x)A_{32}(-x),\quad
g(y)=-A_{31}(y)A_{13}(-y),\quad
h(z)=-A_{12}(z)A_{21}(-z).
$$
Then
$$
\Phi_{21}+\Phi_{13}+\Phi_{32}=0
$$
leads to
$$
A_{23}A_{31}\big( A_{13}A\sp\prime_{32}-A\sp\prime_{13}A_{32}\big)+
A_{31}A_{12}\big( A_{21}A\sp\prime_{13}-A\sp\prime_{21}A_{13}\big)+
A_{12}A_{23}\big( A_{32}A\sp\prime_{21}-A\sp\prime_{32}A_{21}\big)
=0
$$
and consequently
\begin{equation}
g(y)A_{23}A\sp\prime_{32}-f(x)A_{31}A\sp\prime_{13}+
h(z)A_{31}A\sp\prime_{13}-g(y)A_{12}A\sp\prime_{21}+
f(x)A_{12}A\sp\prime_{21}-h(z)A_{23}A\sp\prime_{32}=0.
\label{eq:123}
\end{equation}
Similarly from $\Phi_{12}+\Phi_{23}+\Phi_{31}=0$ we obtain
\begin{equation}
f(x)A_{13}A\sp\prime_{31}-g(y)A_{32}A\sp\prime_{23}+
h(z)A_{32}A\sp\prime_{23}-f(x)A_{21}A\sp\prime_{12}
+g(y)A_{21}A\sp\prime_{12}-h(z)A_{13}A\sp\prime_{31}=0.
\label{eq:132}
\end{equation}
Now
$$
f\sp\prime(x)=A_{23}A\sp\prime_{32}-A_{32}A\sp\prime_{23},\quad
g\sp\prime(y)=A_{31}A\sp\prime_{13}-A_{13}A\sp\prime_{31},\quad
h\sp\prime(z)=A_{12}A\sp\prime_{21}-A_{21}A\sp\prime_{12},
$$
so upon adding (\ref{eq:123}) and (\ref{eq:132}) we obtain
\begin{equation}
g(y)f\sp\prime(x)-f(x)g\sp\prime(y)+h(z)g\sp\prime(y)-
g(y)h\sp\prime(z)+f(x)h\sp\prime(z)-h(z)f\sp\prime(x)=0.
\label{eq:lemt}
\end{equation}
Equation (\ref{eq:lemt}) may be rewritten as
\begin{equation}
\left|\matrix{1&1&1\cr f(x)&g(y)&h(z)\cr
         f\sp\prime(x)&g\sp\prime(y)&h\sp\prime(z)\cr}\right|=0,
\quad\quad x+y+z=0
\label{eq:BP}
\end{equation}
which is the equation studied by Buchstaber and Perelomov.
The solutions to (\ref{eq:BP}) are called nondegenerate if
each of $f(x)$, $g(y)$ and $h(z)$ have poles lying in some finite
domain of the complex plane. 
Degenerate solutions may then be obtained from these.
The nondegenerate solutions to (\ref{eq:BP}) are given by
\begin{equation}
f(x)=\alpha\wp(x-a_1)+\beta,\quad
g(y)=\alpha\wp(y-a_2)+\beta,\quad
h(z)=\alpha\wp(z-a_3)+\beta,
\end{equation}
with $a_1+a_2+a_3=0$.

We now observe that, although we have not yet specified
$A_{12}$, $A_{23}$ and $A_{31}$, we have in fact obtained the
interactions in the present situation. We have
$$
V_{jk}=-\big( A_{jk} A_{kj}\big)\sp\prime
$$
and so 
\begin{equation}
V_{jk}(q_j-q_k)=-\alpha\wp\sp\prime(q_j-q_k+a_j-a_k).
\end{equation}
Thus our functional equation determines the interaction for us. As
for the Lax-pair we may simply choose $A_{jk}(x)=-A_{kj}(-x)$ or
some other form that suits our purpose. Using the addition
properties of the elliptic functions another choice for
$A_{jk}(x)$ could be
\begin{equation}
A_{jk}(x)=\sqrt{\alpha}{\sigma(b-x+\lambda_j-\lambda_k)\over 
\sigma(b)\sigma(x-\lambda_j+\lambda_k)}
\end{equation}
where $\alpha\wp(b)=-\beta$ and $a_i=\lambda_j-\lambda_k$ for
cyclic $i,j,k$.

\section{Discussion}
This paper has introduced a new ansatz (\ref{ansatz})
for a Lax pair describing
systems of particles on the line interacting via pairwise forces (\ref{ans}).
Unlike  existing ansatz we allow these forces to depend in principle
on the particle pair, and so the one ansatz encompasses for example
those of  the Calogero-Moser and Toda systems within a unified framework.
A consequence of allowing varying pairwise interactions is that
the consistency equations for the Lax pair now become a system
of functional equations rather than a single functional equation.
Our approach has been to first study the constituent functional
equations, of interest in their own right, and then to examine
the contraints imposed by the system of which they are a part.

Two quite interesting functional equations 
(\ref{functional}), (\ref{functional2}) arise in this
manner. The first, which arises when
$b\ne0$, has a large group of symmetries acting on it
and we have been able to give its general analytic solution with
appropriate orbits corresponding to the relativistic Calogero-Moser
and Toda interactions. It is this large symmetry group of (\ref{functional})
that enables us to relate previously distinct functional
equations and different physical models.
We remark that a particular case of this
equation has recently arisen in the work \cite{BFV} (see their equation
(13) and Lemma 10) which examines the connection between
functional equations and Dunkl operators.

Unfortunately we have not been able to say as much about the
functional equation (\ref{functional2}) or the associated
system of equations when $b=0$. Certainly the $b\rightarrow0$
limit of our general solution yields a $b=0$ solution, corresponding
to an appropriate nonrelativistic limit, but the nonrelativistic
equations are less rigid.
Similarly we note that both (\ref{functional}), (\ref{functional3}) 
as well as the 
functional equations satisfied by Baker-Akhiezer functions \cite{BK}
are particular cases of (\ref{Bfun}).

The final step in our approach has been 
to examine the contraints imposed by the
system of functional equations on the parameters appearing in the
solutions to (\ref{functional}) and (\ref{functional2}). 
The constraints for the relativistic system were encapsulated in
Theorem $3$.
Although we have a conceptually straightforward unification of
various ansatz for Lax pairs, this stage of our approach  is the
most tedious as it can often involve case by case analysis.
We plan to return to the equation (\ref{eq:relts}) in a later
work.
Finally we have  shown how the examples of 
the known relativistic and nonrelativistic Toda and
Calogero-Moser models arise in our approach as well as introducing a new
system.

{\bf Acknowledgements:} 
One of us (V.M.B.) thanks the Royal Society for a Kapitza Fellowship
in $1993$ and the EPSRC for a Visiting Fellowship.

\appendix
\section{Proof of Theorem 2}
In proving this theorem we shall consider  the constraints imposed by
(\ref{dfd}) and (\ref{dfe}) separately.
In both cases we proceed by first finding relations amongst the constants 
appearing in (\ref{eq:solform}) by equating the poles and zeros
of the various terms given by our first theorem. 
This determines the functions $\psi_{jm}$ and $\psi_{mk}$ up to the action of the
function $f(x)$ and an exponential. By comparing with (\ref{eq:alt}) we then
determine $G_{jm}$ and $G_{mk}$ up to constants.
\subsection{Consistency for (\ref{dfd})}
We begin with (\ref{dfd}).  In this case the theorem says
$$
{A_{jm}(x)/ \psi_{jm}(x)}=f(x)e\sp{-\lambda\sp\prime x}
\big( u_{11}\Phi(x;\nu_{jk})+u_{12}\Phi\sp\prime(x;\nu_{jk})\big)
$$
and
$$
{1/ \psi_{jm}(x)}=f(x)e\sp{\lambda\sp{\prime\prime}x}
\big( v_{11}\Phi(x;\mu_{jk})+
v_{12}\Phi\sp\prime(x;\mu_{jk})\big).
$$
The ratio of these two equations when
$$A_{jm}(x)=c_{jm}{\Phi(x;\nu_{jm})\over\Phi(x;\mu_{jm})} e\sp{\lambda_{jm}x}$$
yields
$$
A_{jm}(x)=c_{jm}{\Phi(x;\nu_{jm})\over\Phi(x;\mu_{jm})} e\sp{\lambda_{jm}x}
={ {u_{11}\Phi(x;\nu_{jk})+u_{12}\Phi\sp\prime(x;\nu_{jk})}\over
   { v_{11}\Phi(x;\mu_{jk})+v_{12}\Phi\sp\prime(x;\mu_{jk})} }
e\sp{\lambda_{jk}x}
$$
\begin{equation}
\phantom{a}\quad\quad
= {u_{12}\over v_{12}}{\Phi(x;\nu_{jk})\over \Phi(x;\mu_{jk})}
{\big[{ u_{11}/u_{12}+\zeta(\nu_{jk})-\zeta(x)-\zeta(\nu_{jk}-x)}\big]\over
 \big[{ v_{11}/v_{12}+\zeta(\mu_{jk})-\zeta(x)-\zeta(\mu_{jk}-x)}\big]}
e\sp{\lambda_{jk}x}.
\label{eq:appa}
\end{equation}
(We will assume $u_{12}$ and $v_{12}$ are nonvanishing and later see that this
is this case. Certainly by considering the behaviour of (\ref{eq:appa}) as $x\rightarrow0$
we see $u_{12}=0\Leftrightarrow v_{12}=0\Leftrightarrow A_{jm}(x)=c\, A_{jk}(x)\,
e\sp{\lambda x}$.)
Now the left hand side has a zero at $\nu_{jm}$ and pole at $\mu_{jm}$.
Equating these with the right hand side (for $\nu_{jm}\ne\nu_{jk}$ and
$\mu_{jm}\ne\mu_{jk}$) shows
$$
u_{11}/u_{12}=\zeta(\nu_{jk}-\nu_{jm})-\zeta(\nu_{jk})+\zeta(\nu_{jm})
$$
$$
v_{11}/v_{12}=\zeta(\mu_{jk}-\mu_{jm})-\zeta(\mu_{jk})+\zeta(\mu_{jm}).
$$
Thus, after making use of (\ref{eq:zetas}),
\begin{equation}
u_{11}/u_{12}+\zeta(\nu_{jk})-\zeta(x)-\zeta(\nu_{jk}-x)=
{\sigma(\nu_{jk})\sigma(x-\nu_{jm})\sigma(x-\nu_{jk}+\nu_{jm}) \over
 \sigma(\nu_{jm})\sigma(\nu_{jk}-\nu_{jm})\sigma(x-\nu_{jk})\sigma(x)}
\label{eq:nus}
\end{equation}
and
\begin{equation}
v_{11}/v_{12}+\zeta(\mu_{jk})-\zeta(x)-\zeta(\mu_{jk}-x)=
{\sigma(\mu_{jk})\sigma(x-\mu_{jm})\sigma(x-\mu_{jk}+\mu_{jm}) \over
 \sigma(\mu_{jm})\sigma(\mu_{jk}-\mu_{jm})\sigma(x-\mu_{jk})\sigma(x)}.
\label{eq:mus}
\end{equation}
Utilising (\ref{eq:nus}) and (\ref{eq:mus}) in (\ref{eq:appa}) now gives
$$
c_{jm}e\sp{(\lambda_{jm}+\zeta(\nu_{jm})-\zeta(\mu_{jm}))x}=
{u_{12}\over v_{12}} 
{\sigma(x-\nu_{jk}+\nu_{jm})\sigma(\mu_{jk}-\mu_{jm})\over
\sigma(x-\mu_{jk}+\mu_{jm})\sigma(\nu_{jk}-\nu_{jm})}
e\sp{(\lambda_{jk}+\zeta(\nu_{jk})-\zeta(\mu_{jk}))x}
$$
from which we deduce
\begin{eqnarray}
\nu_{jk}-\nu_{jm}&=&\mu_{jk}-\mu_{jm}\hfill\hfill\phantom{a}
\label{eq:nu1}\\
\lambda_{jm}+\zeta(\nu_{jm})-\zeta(\mu_{jm})&=&
\lambda_{jk}+\zeta(\nu_{jk})-\zeta(\mu_{jk})\label{eq:appls}\\
c_{jm}v_{12}&=&u_{12}.\hfill\hfill\phantom{a}
\end{eqnarray}
>From (\ref{eq:appls}) and our expressions for $u_{11}\over u_{12}$ and
$v_{11}\over v_{12}$ we find
\begin{equation}
\lambda_{jm}-\lambda_{jk}={v_{11}\over v_{12}}-{u_{11}\over u_{12}}.
\end{equation}

The same considerations now applied to ${A_{mk}(x)/ \psi_{mk}(x)}$
rather than ${A_{jm}(x)/ \psi_{jm}(x)}$ similarly show
\begin{eqnarray}
u_{21}/u_{22}&=&\zeta(\nu_{jk}-\nu_{mk})-\zeta(\nu_{jk})+\zeta(\nu_{mk})
\nonumber\\
v_{21}/v_{22}&=&\zeta(\mu_{jk}-\mu_{mk})-\zeta(\mu_{jk})+\zeta(\mu_{mk})
\nonumber\\
\nu_{jk}-\nu_{mk}&=&\mu_{jk}-\mu_{mk}\hfill\hfill\phantom{a}\\
\lambda_{mk}+\zeta(\nu_{mk})-\zeta(\mu_{mk})&=&
\lambda_{jk}+\zeta(\nu_{jk})-\zeta(\mu_{jk})\\
c_{mk}v_{22}&=&u_{22}\hfill\hfill\phantom{a}\\
\lambda_{mk}-\lambda_{jk}&=&{v_{21}\over v_{22}}-{u_{21}\over u_{22}}.
\hfill\hfill
\end{eqnarray}
Combining these relations with (\ref{eq:nu1}) and (\ref{eq:appls})
thus proves (\ref{eq:relmns}) and (\ref{eq:rellmns}) for the case
being examined.

We have yet to impose the constraint  $\det{U}=c_{jk}\det{V}$. Now
$$
c_{jk}={\det{U}\over\det{V}}= {u_{12} u_{22}\over v_{12}v_{22} }
{\big( {u_{11}\over u_{12}}-{u_{21}\over u_{22}}\big)\over
 \big( {v_{11}\over v_{12}}-{v_{21}\over v_{22}}\big)}
$$
$$
=c_{jm}c_{mk}{\big(\zeta(\nu_{jk}-\nu_{jm})-\zeta(\nu_{jk})+\zeta(\nu_{jm})
\big)-\big(
\zeta(\nu_{jk}-\nu_{mk})-\zeta(\nu_{jk})+\zeta(\nu_{mk}) \big)\over
\big(\zeta(\mu_{jk}-\mu_{jm})-\zeta(\mu_{jk})+\zeta(\mu_{jm})\big)-\big(
\zeta(\mu_{jk}-\mu_{mk})-\zeta(\mu_{jk})+\zeta(\mu_{mk})\big)}
$$
from which (\ref{eq:relts}) follows. Substituting these results 
immediately yields (\ref{eq:reltas}).

After simplifying and again using
(\ref{eq:zetas}) we obtain
\begin{equation}
{c_{jk}\over c_{jm}c_{mk}}=
{\sigma(\nu_{jk})\over \sigma(\mu_{jk})}
{\sigma(\mu_{jm})\over \sigma(\nu_{jm})}
{\sigma(\mu_{mk})\over \sigma(\nu_{mk})}
{\sigma(\nu_{jm}+\nu_{mk}-\nu_{jk})\over \sigma(\mu_{jm}+\mu_{mk}-\mu_{jk})}.
\label{eq:appacs}
\end{equation}

At this stage we have obtained
$$
1/\psi_{jm}(x)=-v_{12}\,f(x) e\sp{[\lambda\sp{\prime\prime}+\zeta(\mu_{jk})]x}
{\sigma(x-\mu_{jm})\sigma(x-\mu_{jk}+\mu_{jm})\over
\sigma\sp2(x) \sigma(\mu_{jm})\sigma(\mu_{jk}-\mu_{jm})}
$$
with a similar expression holding for $1/\psi_{mk}$ and the desired expressions
for $A_{jm}$, $A_{mk}$ and $A_{jk}$.

The final constraints arise by considering (\ref{eq:alt}) which may be recast as
$$
G_{jm}(x)-G_{mk}(y)=
\partial \ln\left(1 -\frac{A_{jm}(x) A_{mk}(y)}{A_{jk}(x+y)}\right).
$$
Now each side of this equation may be separately calculated and on comparison we find our 
last constraint.
Using the form of $A_{jm}$, $A_{mk}$ and $A_{jk}$ given by (\ref{eq:reltas})
we obtain for the left hand side
$$
\zeta(x+\mu_{mk}-\mu_{jk}) -\zeta(x-\mu_{jm}) -\zeta(y+\mu_{jm}-\mu_{jk})+
 \zeta(y-\mu_{mk}),
$$
where substantial use has been made of (\ref{eq:zetas}).
On the other hand, determining $G_{jm}(x)$  directly from our expressions for
$\psi_{jm}$  yields
$$G_{jm}(x)= -\zeta(x-\mu_{jm})-\zeta(x+\mu_{jm}-\mu_{jk})-F(x),$$
where $F(x)$ encodes the remaining functional dependence of $\psi_{jm}$. 
Comparison shows
$$F(x)=-\zeta(x+\mu_{jm}-\mu_{jk})-\zeta(x+\mu_{mk}-\mu_{jk}) +const$$
and so we  have the final relation (\ref{eq:finrels})
$$ G_{jm}(x)= -\zeta(x-\mu_{jm})+\zeta(x+\mu_{mk}-\mu_{jk}) +const.$$
We may for example choose
$$
f(x)=-{\sigma(\mu_{jk}-\mu_{jm})\sigma(\mu_{jk}-\mu_{mk})\sigma\sp2(x)\over
\sigma(x-\mu_{jk}+\mu_{jm})\sigma(x-\mu_{jk}+\mu_{mk})}.
$$
Still $f(x)$ is only determined up to a constant multiple of an exponential.
This gives
$$
1/\psi_{jm}(x)=v_{12} {\sigma(x-\mu_{jm})\sigma(\mu_{jm}-\bar\mu)\over
 \sigma(\mu_{jm})\sigma(x+\bar\mu-\mu_{jm})} e\sp{\kappa x}
$$
where we have set $\bar\mu=-\mu_{jk}+\mu_{jm}+\mu_{mk}$ and $\kappa $ is
an arbitrary constant. A similar
expression holds for $1/\psi_{mk}$.

\subsection{Consistency for (\ref{dfe})}
We next consider the constraints arising from the consistency of $(22)$
when again $A_{jk}$, $A_{jm}$ and $A_{mk}$ are given by (\ref{eq:solform})
but  now $A_{jm}(x)=c_2 A_{mk}(x)$. The latter of course means
$\nu_{jm}=\nu_{mk}$, $\mu_{jm}=\mu_{mk}$, $\lambda_{jm}=\lambda_{mk}$
and $c_{jm}=c_2 c_{mk}$. 
We now must find a function $f(x)$,  matrices $U,V$  and $\lambda\sp\prime$, 
$\lambda\sp{\prime\prime}$ such that
$$
\pmatrix{ {c_2 A_{mk}(x)}\cr {C_{mk}(x)} \cr}
=f(x)e\sp{-\lambda\sp\prime x} U
\pmatrix{ \Phi(x,\nu_{jk})\cr \Phi\sp\prime(x,\nu_{jk})\cr},
$$
$$
\pmatrix{{G_{mk}(x)}\cr1\cr}
=f(x)e\sp{\lambda\sp{\prime\prime}x} V
\pmatrix{ \Phi(x,\mu_{jk})\cr \Phi\sp\prime(x,\mu_{jk})\cr}.
$$
We proceed in much the same manner as in the previous case and accordingly
we will be less detailed. Now
\begin{eqnarray*}
{c_2 A_{mk}(x)\over 1}&=&e\sp{-(\lambda\sp\prime+\lambda\sp{\prime\prime})x}
{ {u_{11}\Phi(x;\nu_{jk})+u_{12}\Phi\sp\prime(x;\nu_{jk})}\over
   { v_{21}\Phi(x;\mu_{jk})+v_{22}\Phi\sp\prime(x;\mu_{jk})} }\\
&=&
 {u_{12}\over v_{22}}{\Phi(x;\nu_{jk})\over \Phi(x;\mu_{jk})}
{\big[{ u_{11}/u_{12}+\zeta(\nu_{jk})-\zeta(x)-\zeta(\nu_{jk}-x)}\big]\over
 \big[{ v_{21}/v_{22}+\zeta(\mu_{jk})-\zeta(x)-\zeta(\mu_{jk}-x)}\big]}
e\sp{\lambda_{jk}x}.
\end{eqnarray*}
Again a comparison of zeros and poles shows
\begin{eqnarray*}
u_{11}/u_{12}&=&\zeta(\nu_{jk}-\nu_{mk})-\zeta(\nu_{jk})+\zeta(\nu_{mk})\\
v_{21}/v_{22}&=&\zeta(\mu_{jk}-\mu_{mk})-\zeta(\mu_{jk})+\zeta(\mu_{mk})\\
\nu_{jk}-\nu_{mk}&=&\mu_{jk}-\mu_{mk}\hfill\hfill\phantom{a}\\
\lambda_{mk}+\zeta(\nu_{mk})-\zeta(\mu_{mk})&=&
\lambda_{jk}+\zeta(\nu_{jk})-\zeta(\mu_{jk})\\
c_2 \, c_{mk}v_{22}&=&u_{12}\hfill\hfill\phantom{a}\\
\lambda_{mk}-\lambda_{jk}&=&{v_{21}\over v_{22}}-{u_{21}\over u_{22}}.
\hfill\hfill
\end{eqnarray*}

These are the exact analogues of our earlier equations. In addition we have
using the definition of $C$ that
$$
{C_{mk}\over  A_{mk}}=c_2
{ {u_{21}\Phi(x;\nu_{jk})+u_{22}\Phi\sp\prime(x;\nu_{jk})}\over
   { u_{11}\Phi(x;\nu_{jk})+u_{12}\Phi\sp\prime(x;\nu_{jk})} }
={A\sp\prime_{mk}\over A_{mk}}-{G_{mk}}
$$
as well as
$$
{G_{mk}\over1}={{ v_{11}\Phi(x;\mu_{jk})+v_{12}\Phi\sp\prime(x;\mu_{jk})}
\over{ v_{21}\Phi(x;\mu_{jk})+v_{22}\Phi\sp\prime(x;\mu_{jk})} }.
$$
Upon using substituting the form of $A_{mk}$ theses equations may be
rearranged to give
$$
c_2 { {u_{21}+u_{22}[\zeta(\nu_{jk})-\zeta(x)-\zeta(\nu_{jk}-x)]}\over
       {u_{11}+u_{12}[\zeta(\nu_{jk})-\zeta(x)-\zeta(\nu_{jk}-x)]}}
+{ {v_{11}+v_{12}[\zeta(\mu_{jk})-\zeta(x)-\zeta(\mu_{jk}-x)]}\over
       {v_{21}+v_{22}[\zeta(\mu_{jk})-\zeta(x)-\zeta(\mu_{jk}-x)]}}
-\lambda_{mk}\nonumber
$$
\begin{equation}
=
{\sigma(x)\sigma(\mu_{mk}-\nu_{mk})\sigma(\mu_{mk}+\nu_{mk}-x) \over
\sigma(\nu_{mk})\sigma(x-\nu_{mk})\sigma(\mu_{mk})\sigma(\mu_{mk}-x)}.
\label{eq:appba}
\end{equation}
As $x\rightarrow0$ we see
$$
c_2 {u_{22}\over u_{12}}+{v_{12}\over v_{22}}-\lambda_{mk}=0.
$$
Upon making use of this and further simplifying (\ref{eq:appba}) yields
$$
c_2 {u_{22}\over u_{12}}\big( {u_{21}\over u_{22}}+\zeta(\nu_{jk})
-\zeta(\nu_{mk})-\zeta(\nu_{jk}-\nu_{mk})\big)
{ \sigma(\nu_{mk})\sigma(\nu_{jk}-\nu_{mk})\sigma(x-\nu_{jk})\over
\sigma(\nu_{jk}) \sigma(x-\nu_{mk})\sigma(x+\nu_{mk}-\nu_{jk})}
$$
$$
+
{v_{12}\over v_{22}}\big( {v_{11}\over v_{12}}+\zeta(\mu_{jk})
-\zeta(\mu_{mk})-\zeta(\mu_{jk}-\mu_{mk})\big)
{ \sigma(\mu_{mk})\sigma(\mu_{jk}-\mu_{mk})\sigma(x-\mu_{jk})\over
\sigma(\mu_{jk}) \sigma(x-\mu_{mk})\sigma(x+\mu_{mk}-\mu_{jk})}
$$
$$
=
{\sigma(\mu_{mk}-\nu_{mk})\sigma(\mu_{mk}+\nu_{mk}-x)\over
\sigma(\nu_{mk})\sigma(x-\nu_{mk})\sigma(\mu_{mk}) \sigma(x-\mu_{mk})}
$$
As $x\rightarrow \mu_{jk}$ we find
$$
c_2 {u_{22}\over u_{12}}\big({ u_{21}\over u_{22}}- { u_{11}\over u_{12}}
\big)
{\sigma\sp2(\nu_{mk})\sigma\sp2(\nu_{jk}-\nu_{mk})\over
\sigma(\nu_{jk})\sigma(2\nu_{mk}-\nu_{jk})}
=-1
$$
and as $x\rightarrow \nu_{jk}$ we obtain
$$
{v_{12}\over v_{22}}\big( {v_{11}\over v_{12}}-{v_{21}\over v_{22}}\big)
{\sigma\sp2(\mu_{mk})\sigma\sp2(\mu_{jk}-\mu_{mk})\over
\sigma(\mu_{jk})\sigma(2\mu_{mk}-\mu_{jk})}
=1.
$$
No further constraints are imposed on the matrices $U$ and $V$.
The ratio of these last two equations (taking into account that
$\det{U}=c_{jk}\det{V}$) then yields
$$
{c_{jk}\over c_2\, c\sp2_{mk}}=
{\sigma(\nu_{jk})\over\sigma(\mu_{jk})}
{\sigma\sp2(\mu_{mk})\over\sigma\sp2(\nu_{mk})}
{\sigma(2\nu_{mk}-\nu_{jk})\over \sigma(2\mu_{mk}-\mu_{jk})}
$$
which is just the $\nu_{jm}\rightarrow \nu_{mk}$ limit of our previous
result and we again obtain the relations stated in the theorem by similar
analysis.

\section{Proof of Lemma 4}
Differentiating (\ref{eqlm4f}) with respect to $y$ yields
$$
F\sp\prime(x+y)=\phi(x)\psi\sp{\prime\prime}(y)-\phi\sp\prime(x)\psi\sp\prime(y),
$$
and so upon letting $y=0$ we must solve the two equations
\begin{equation}
F(x)=\phi(x)\psi\sp\prime(0)-\phi\sp\prime(x)\psi(0),
\label{proofa}
\end{equation}
\begin{equation}
F\sp\prime(x)=\phi(x)\psi\sp{\prime\prime}(0)-\phi\sp\prime(x)\psi\sp\prime(0).
\label{proofb}
\end{equation}
Now either $F(x)\not\equiv 0$ or
$F(x)\equiv 0$ and $\phi(x)=c_1\exp(\lambda x)$,
and $\psi(x)=c_2\exp(\lambda x)$.
In the former case (perhaps by translating $y$ if necessary)
either $\psi(0)\ne0$ or $\psi\sp\prime(0)\ne0$
and we consider these two cases separately.

If $\psi(0)\ne0$ then utilising
the group of symmetries of the functional equation we may 
set $\psi(0)=1$ and $\psi\sp\prime(0)=0$. From
(\ref{proofa}) and (\ref{proofb})
$$
F(x)=-\phi\sp\prime(x)\quad\quad 
F\sp\prime(x)=\psi\sp{\prime\prime}(0)\phi(x)
$$
and so
$$\phi\sp{\prime\prime}(x)=\lambda\sp2\phi(x)
$$
where $\lambda\sp2=-\psi\sp{\prime\prime}(0)$. Thus
$\phi(x)=\tilde c_1\exp(\lambda x)+\tilde c_2\exp(-\lambda x)$ and 
$F(x)=-\lambda (\tilde c_1\exp(\lambda x)-\tilde c_2\exp(-\lambda x))$.
Therefore
\begin{eqnarray*}
F(x+y)&=& -\lambda \tilde c_1\exp\lambda( x+y)+\lambda \tilde c_2\exp-
\lambda( x+y)\\
&=&\tilde c_1\exp(\lambda x)\big( \psi\sp\prime(y)-\lambda\psi(y)\big)
+\tilde c_2\exp(-\lambda x)\big( \psi\sp\prime(y)+\lambda\psi(y)\big)
\end{eqnarray*}
which upon rearranging yields
$$
0=\tilde c_1\exp(\lambda x)\big( \psi\sp\prime(y)-\lambda\psi(y)+\lambda
\exp(\lambda y)\big)
+\tilde c_2\exp(-\lambda x)\big( \psi\sp\prime(y)+\lambda\psi(y)-\lambda
\exp(\lambda y)\big).
$$
Thus either $\tilde c_1=0$ or $\tilde c_2=0$. The latter leads to
$$
0=\psi\sp\prime(y)-\lambda\psi(y)+\lambda
\exp(\lambda y)
$$
and consequently (with the chosen initial conditions)
\begin{equation}
\psi(y)=(1-\lambda y)\exp(\lambda y).
\end{equation}
which is of the required form.

Finally, if $\psi(0)=0$, we may set $\psi\sp\prime(0)=1$
and $\psi\sp{\prime\prime}(0)=0$. Now we have
$F(x)=\phi(x)$ and $F\sp\prime(x)=-\phi\sp\prime(x)$,
whence $\phi(x)=c_1$ is a constant.
In this case
\begin{equation}
\psi(y)=c_2 +y,
\end{equation}
again of the required form.

{\hfill $\square$}

\end{document}